\documentclass[11pt]{article}
\usepackage[a4paper,hmarginratio=1:1,vmarginratio=2:3,
totalwidth=15.6cm,totalheight=22.39cm]{geometry}
\usepackage{bm,epstopdf,epsfig,amsmath,amssymb,amsfonts,colordvi,latexsym,comment,cancel,
verbatim}
\usepackage{graphicx,graphics}
\usepackage[font=md,captionskip=8pt]{subfig}
\usepackage[usenames,dvipsnames]{color}
\def\AdS{\mathrm{AdS}}

\usepackage{latexsym,amsmath,amssymb,amscd,cite}
\usepackage{pdfsync}
\usepackage{epsf}
\usepackage[all]{xy}
\usepackage{amsthm}

\DeclareSymbolFont{extraup}{U}{zavm}{m}{n}
\DeclareMathSymbol{\varheart}{\mathalpha}{extraup}{86}
\DeclareMathSymbol{\vardiamond}{\mathalpha}{extraup}{87}

\newcommand{\cinf}{{{\cal \cC}^\infty(M,\R)}}

\newcommand{\bcO}{\boldsymbol{\check{\cal O}}}

\newtheorem{Theorem}{Theorem}[section]
\newtheorem{Lemma}[Theorem]{Lemma}
\newtheorem{Proposition}[Theorem]{Proposition}
\theoremstyle{definition}
\newtheorem{Definition}[Theorem]{Definition}
\theoremstyle{remark}

\def\ad{\mathrm{ad}}

\newcommand{\bp}{\begin{Proposition}}
\newcommand{\ep}{\end{Proposition}}
\newcommand{\bl}{\begin{Lemma}}
\newcommand{\el}{\end{Lemma}}
\newcommand{\bt}{\begin{Theorem}}
\newcommand{\et}{\end{Theorem}}
\newcommand{\bd}{\begin{Definition}}
\newcommand{\ed}{\end{Definition}}
\newcommand{\End}{\mathrm{End}}

\newcommand{\eqdef}{\stackrel{{\rm def.}}{=}}

\DeclareFontFamily{U}{rsf}{}
\DeclareFontShape{U}{rsf}{m}{n}{<5> <6> rsfs5 <7> <8> <9> rsfs7 <10-> rsfs10}{}
\DeclareMathAlphabet\Scr{U}{rsf}{m}{n}

\newcommand{\KA}{K\"{a}hler-Atiyah~}

\def\R{{\Bbb R}}

\def\rk{{\rm rk}}

\def\dd{\mathrm{d}}

\def\vol{\mathrm{vol}}
\def\AdS{\mathrm{AdS}}

\newcommand{\be}{\begin{equation*}}
\newcommand{\ee}{\end{equation*}}
\newcommand{\ben}{\begin{equation}}
\newcommand{\een}{\end{equation}}
\newcommand{\beqa}{\begin{eqnarray*}}
\newcommand{\eeqa}{\end{eqnarray*}}
\newcommand{\beqan}{\begin{eqnarray}}
\newcommand{\eeqan}{\end{eqnarray}}

\newcommand{\nn}{\nonumber}

\newcommand{\tr}{\mathrm{tr}}

\def\cC{{\mathcal C}}
\def\cB{\Scr B}

\def\cK{\mathrm{\cal K}}

\def\Spin{\mathrm{Spin}}

\def\cD{\mathcal{D}}

\def\cC{\mathcal{C}}
\def\SU{\mathrm{SU}}

\def\G_2{\mathrm{G_2}}

\def\cS{\mathcal{S}}

\def\btu{\bigtriangleup}

\def\cX{\mathcal{X}}

\newcommand{\twopartdef}[4]
{
	\left\{
		\begin{array}{ll}
			#1 & \mbox{if } #2 \\
			#3 & \mbox{if } #4
		\end{array}
	\right.
}

\usepackage{fancyhdr}
\begin{document}
\thispagestyle{empty}

 \vspace{3.7cm}

\begin{center}

\Large{\bf{A generalization of  Calabi-Yau fourfolds arising from M-theory compactifications}}

\vspace{1cm}

\large{Elena Mirela Babalic$^{1,2}$, Calin Iuliu Lazaroiu$^3$ }

\vspace{0.5cm}

\small{$^1$Department of Physics, University of Craiova, 13 Al.~I. Cuza Str., Craiova 200585, Romania \\

$^2$Department of Theoretical Physics, National Institute of Physics and Nuclear Engineering,\\
Str. Reactorului no.30, P.O.BOX MG-6, Bucharest-Magurele  077125,  Romania \\

$^3$Center for Geometry and Physics, Institute for Basic Science (IBS), 77 Cheongam-ro,\\
 Nam-gu, Pohang, Gyeongbuk, Korea 790-784 }

\vspace{0.3cm}

\texttt{mbabalic@theory.nipne.ro, calin@ibs.re.kr}

\end{center}

\vspace{0.5cm}

\begin{abstract}
\noindent Using a reconstruction theorem, we prove that the supersymmetry conditions for a certain class of 
flux backgrounds are equivalent with a tractable subsystem of relations on differential forms which 
encodes the full set of contraints arising fom Fierz identities and from the differential and algebraic 
conditions on the internal part of the supersymmetry generators. 
The result makes use of the formulation  of such problems through K\"{a}hler-Atiyah bundles, which we
developed in previous work. Applying this to the most general ${\cal N}=2$ flux compactifications 
of 11-dimensional supergravity on 8-manifolds, we can extract the conditions constraining such 
backgrounds and give  an overview of the resulting geometry, which generalizes that of Calabi-Yau fourfolds.
 
\

\end{abstract}

\section[]{Introduction}

The geometry of the most general ${\cal N}=2$ compactifications of $M$-theory on eight-manifolds is of interest, for example, 
for resolving certain issues in $F$-theory. The analysis of this class of backgrounds was initiated in our 
previous papers using the methods of geometric algebra, which prove to be extremely useful computationally. We give a brief 
announcement of new results regarding this class of compactifications.  

After recally some basic facts from \cite{ga1, MartelliSparks, Tsimpis} 
for the case ${\cal N}=1$ and from \cite{ga2, Palti} for the case ${\cal N}=2$, we give a reconstruction theorem 
for eight-domensional Majorana spinors within the frame of geometric algebra, emphasizing its use for encoding the full set of 
contraints arising fom Fierz identities and from the differential and algebraic conditions on the internal part of the supersymmetry 
generators into constraints on differential forms, for any amount of supersymmetry preserved by the effective action. 
We exemplify this for ${\cal N}=1$ and ${\cal N}=2$. We also briefly announce some results regarding the geometry 
of such backgrounds. Our notations and conventions are explained in \cite{ga1}.

\section{Introduction to  compactifications of M-theory to $\AdS_3$}
\label{sec:2}

Consider M-theory warped compactifications down to $\AdS_3$ (or Minkowski) spaces with cosmological constant 
$\Lambda=-8\kappa^2$, where $\kappa\geq 0$ (with $\kappa=0$ giving the Minkowski limit). The 11-dimensional 
background is of the form ${\tilde M}=N\times M$, where the external space $N$ is an oriented 3-manifold 
diffeomorphic with $\R^3$ and carrying the $\AdS_3$ metric, while the internal space $M$ is an oriented spinnable 
Riemannian 8-manifold with metric $g$. The metric on ${\tilde M}$, denoted by $\tilde g$, is a warped product:
\beqa
\label{warpedprod}
\dd {\tilde s}^2_{11}  & = & e^{2\Delta} \dd s_{11}^2~~~{\rm where}~~~\dd s_{11}^2=\dd s^2_3+ g_{mn} \dd x^m \dd x^n~~,
\eeqa
where the warp factor $\Delta$ is a smooth function defined on $M$ and $\dd s_3^2$ is the squared length element on $N$. 

The supergravity action in 11 dimension involves the metric $\tilde g$, the 3-form potential $\tilde C$ and the gravitino $\tilde \Psi$ : 
    \be
  S_{11}=\frac{1}{2}\int d^{11} y \left[\tilde R\star 1 - \frac{1}{2}\tilde G\wedge\star\tilde G-
  \frac{1}{6}\tilde G\wedge\tilde G\wedge\tilde C \right] + \mathrm{terms~involving}~{\tilde \Psi}~.\nn
  \ee
For the field strength ${\tilde G}$, the following compactification ansatz is used:
\be
{\tilde G} = e^{3\Delta} G~~~{\rm with}~~~ G = {\rm vol}_3\wedge f+F~~,
\ee
where $\vol_3$ is the volume form of $N$, while the 1-form $f$ and the 4-form $F$ are the fluxes on $M$. 

The supersymmetric flux backgrounts of interest arise when the gravitino and its supersymmetry variation vanish:
\ben
\label{susycond}
\delta_{\tilde\eta}\tilde\Psi_A=\tilde D_A \tilde\eta=0~~,~~A=0,...,10~~, 
\een
where $\tilde D_A $ is the supercovariant connection \cite{ga1, MartelliSparks}. 
For the supersymmetry generator ${\tilde \eta}$, which is a Majorana spinor, one uses the ansatz:
\be
{\tilde \eta}=e^{\frac{\Delta}{2}}\eta~~~{\rm with}~~~ \eta=\varphi\otimes \xi~~,
\ee
where $\xi$ and $\varphi$ are Majorana spinors on $M$ and $N$ respectively.

Assuming that $\varphi$ is a Killing spinor on the $\AdS_3$ space, the supersymmetry condition \eqref{susycond} 
decomposes into a set of constraints for the internal supersymmetry generator $\xi$ 
which we call the constrained generalized Killing (CGK) spinor equations:
\ben
\label{CGK}
D_m\xi=Q\xi=0~~.
\een
Here, $D_m$ is a certain linear connection on $S$ and ${Q}\in \Gamma(M,\End(S))$ is a globally-defined endomorphism 
of the vector bundle $S$. As in \cite{MartelliSparks, Tsimpis}, {\em we do not require that $\xi$ has definite 
chirality} (i.e. $\xi$ need {\em not} satisfy the Weyl condition).  The expressions for $D_m$ and $Q$
 depend on the metric and fluxes and can be found explicitely in the literature, particularly in \cite{ga1}.

The space of solutions of \eqref{CGK} is a finite-dimensional $\R$-linear
 subspace $\cK(D,{Q})$ of the space $\Gamma(M,S)$ of smooth sections of
$S$. Of course, this subspace is trivial for generic metrics $g$ and fluxes $F$ and $f$
 on $M$, since the generic compactification of the type we consider breaks all supersymmetry. 
The interesting problem is to find those metrics and fluxes on $M$ for which some fixed amount of
supersymmetry is preserved in three dimensions, i.e.  for which the
space $\cK(D,{Q})$ has some given non-vanishing dimension, which
we denote by $s$. The case $s=1$ (which corresponds to ${\cal N}=1$
supersymmetry in three dimensions) was studied in
\cite{MartelliSparks, Tsimpis} is recalled below in the context
of the the reconstruction theorem. An important observation made in \cite{ga1,ga2} 
is that the pairing $\cB$ on $S$ is $D_m$-flat:
\ben
\label{flatness}
\partial_m\cB(\xi,\xi')=\cB(D_m\xi,\xi')+\cB(\xi,D_m\xi')~~,~~\forall \xi,\xi'\in\Gamma(M,S)~,
\een
so that $D_m$ is an $O(16)$-connection. 
It can be proved that requiring our background to preserve $s$ independent supersymmetry generators
(i.e. requiring that $\dim \cK(D,{Q})=s$) is {\em equivalent}
to requiring that the system \eqref{CGK} admits $s$ solutions
$\xi_1,\ldots,\xi_s$ which are $\cB$-orthonormal at every point of
$M$.

What we want to do is to solve the system of differential and algebraic constraints resulting from 
\eqref{CGK} for ${\cal N}$=2 (or higher), together with all the Fierz identities involved,
with the purpose of finding all the information about the geometry of the background space. 
Before we do that, let us introduce a simplifying method to what we have previously used in \cite{ga1, ga2}.

\section{A reconstruction theorem for sections of a vector bundle}
\label{sec:3}

\noindent {\bf Notations and conventions.} Here we take $M$ to denote an oriented, 
connected and paracompact smooth manifold of general dimension $d$, whose unital 
$\R$-algebra of smooth real-valued functions we denote by $\cinf$. 
All vector bundles over $M$ are assumed to be smooth.

Let $S$ be a smooth vector bundle of rank $r$ over $M$, endowed with a
scalar product $\cB$. We are interested in describing collections of smooth
global sections $\xi_i\in \Gamma(M,S)$ (where $i=1\ldots s$) through
certain smooth global sections $E_{ij}\eqdef E_{\xi_i,\xi_j}\in
\Gamma(M,\End(S))$ of the bundle of endomorphisms of $S$, which are
built as bilinear combinations of the sections $\xi_i$. More
precisely, we consider the global endomorphisms of $S$ whose action on
smooth sections is given by:  
\ben
\label{Eij}
E_{ij}(\xi)\eqdef \cB(\xi,\xi_j)\xi_i~~,~~\forall \xi\in \Gamma(M,S)~~,~~\forall i,j=1\ldots s~~
\een
and seek a set of conditions satisfied by these objects which allow us to reconstruct the sections $\xi_i$ from knowledge of $E_{ij}$. 
Setting $\cB_{ij}\eqdef \cB(\xi_i,\xi_j)\in \cinf$, an easy computation shows that $E_{ij}$ satisfy: 
\ben
\label{TraceRel}
\tr E_{ij}=\cB_{ij}
\een
 and that ${\cal O}_{ij}=E_{ij}$ is a particular solution of the system of equations:
\ben
\label{S}
\begin{split}
& {\cal O}_{ij}\circ {\cal O}_{kl}=\tr({\cal O}_{kj}) {\cal O}_{il}~~,~~\forall i,j,k,l=1\ldots s\\
& {\cal O}_{ij}^t ={\cal O}_{ji}~~,~~\forall i,j=1\ldots s~~,
\end{split}
\een
where $(~)^t$ denotes the transpose taken with respect to $\cB$. The equations obtained from the first relations of \eqref{S} 
by setting $j=k=l=i$ give $s$ decoupled systems of the same type satisfied by ${\cal O}_i\eqdef {\cal O}_{ii}$:
\ben
\label{diagonal}
\boxed{
\begin{split}
& {\cal O}_i^2 =\tr({\cal O}_i) {\cal O}_i~~\\
& {\cal O}_i^t= {\cal O}_i~~.
\end{split}
}
\een
A particular case of interest in the applications of this paper is when $\xi_i$ are mutually orthonormal at every point of $M$, 
which amounts to setting $\cB_{ij}=\delta_{ij} 1_M$ in \eqref{TraceRel}. 
This leads us to consider the further conditions:
\be
\tr({\cal O}_i)=1_M~~,~~
\ee
the diagonal part of which has to be considered together with \eqref{diagonal}. The one can prove the following: 

\

\noindent {\bf Proposition.} Giving an everywhere orthonormal system of global smooth sections 
$\xi_i\in \Gamma(M,S)$ is equivalent to giving a system
of global endomorphisms ${\cal O}_i\in \Gamma(M,\End(S))$ which satisfy the following conditions:
\ben
\label{ndiagonal}
\boxed{
\begin{split}
& {\cal O}_i^2 = {\cal O}_i~~,\\
& {\cal O}_i^t= {\cal O}_i~~,\\
& \tr({\cal O}_i)=1_M~~,\\
& \tr({\cal O}_i\circ {\cal O}_j)=0_M~~\mathrm{for}~~ i<j~~.
\end{split}
}~
\een
Furthermore, a solution $({\cal O}_i)_{i=1\ldots s}$ of this system determines 
the corresponding everywhere $\cB$-orthonormal system of sections of $S$ via the relations: 
\be
{\cal O}_i=E_{\xi_i,\xi_i}~~,~~\mathrm{where}~~E_{\xi_i,\xi_i}(\xi)\eqdef \cB(\xi,\xi_i)\xi_i~~,
\ee
up to {\em independent} ambiguities of the form $\xi_i\rightarrow -\xi_i$. 

In this approach, the third row equations in \eqref{ndiagonal} impose the unit norm conditions 
$\cB(\xi_i,\xi_i)=1_M$ while the last row of equations impose orthogonality of these sections at all points of $M$.

\paragraph{Constrained flat sections.}

Let $Q\in \Gamma(M,\End(S))$ be a smooth global endomorphism of $S$ and $D:\Gamma(M,S)\rightarrow\Omega^1(M)\otimes_\cinf\Gamma(M,\End(S))$ be a connection 
on $S$ which is compatible with $\cB$ in the sense that $\cB$ is $D$-flat.

By definition, a {\em $Q$-constrained and $D$-flat section} of $S$ is a smooth global section 
$\xi\in \Gamma(M,S)$ which satisfies the CGK conditions \eqref{CGK}. 
Equation \eqref{flatness} for $D$-flatness implies that the $\cB$-pairing of 
any two $D_m$-flat spinors is constant on $M$ and in
particular that any two solutions $\xi_1,\xi_2$ of \eqref{CGK} have
constant $\cB$-pairing.  

Since the parallel transport of $D_m$
preserves $\cB$ by virtue of \eqref{flatness}, it immediately follows
(see \cite{ga1}) that any two solutions of \eqref{CGK} which are
linearly independent at a point are linearly independent everywhere and can be
replaced by two solutions of \eqref{CGK} which are $\cB$-orthogonal
everywhere and whose local values at any point span the same subspace
of the fiber of $S$ at that point as the two original solutions.  This
implies that, when the $\R$-vector space of solutions of \eqref{CGK} is non-vanishing, we can always
find a basis of solutions which consists of sections of $S$ that
are everywhere $\cB$-orthogonal. In \cite{ga1}, we showed that the global endomorphisms 
$E_{\xi,\xi'}$ defined through:
\be
\label{E}
E_{\xi,\xi'}(\xi'')\eqdef \cB(\xi'',\xi')\xi~~,~~\forall \xi''\in \Gamma(M,S)~~
\ee
satisfy the identities: 
\be
\label{DE}
D_m^\ad(E_{\xi,\xi'})=E_{D_m\xi,\xi'}+E_{\xi, D_m\xi'}~~,~~\forall \xi,\xi'\in \Gamma(M,S)~~
\ee
and:
\be
\label{EE}
Q\circ E_{\xi,\xi'}=E_{Q\xi,\xi'}~~,~~E_{\xi,\xi'}\circ Q^t=E_{\xi, Q\xi'}~~,~~\forall \xi,\xi'\in \Gamma(M,S)~~,
\ee
where $D^\ad$ is the connection induced by $D$ on the bundle $\End(S)$. 

\paragraph{Proposition.} Let $\xi\in \Gamma(M,S)$. Then the following statements are equivalent:

\begin{itemize}
\item $Q\xi=0$
\item $Q\circ E_{\xi,\xi}=0$
\item $E_{\xi,\xi}\circ Q^t=0$~~
\end{itemize}

\paragraph{Proposition.} Let $\xi\in \Gamma(M,S)$ such that $\xi$ is nowhere vanishing. Then the following statements are equivalent:

\begin{itemize}
\item $D_m\xi=0$
\item $D_m^\ad(E_{\xi,\xi})=0$
\end{itemize}

\paragraph{Corollary.} Let $\xi\in \Gamma(M,S)$ such that $\xi$ is nowhere vanishing. Then $\xi$ satisfies \eqref{CGK} iff. $E_{\xi,\xi}$ satisfies: 
\be
D_m^\ad(E_{\xi,\xi})=Q\circ E_{\xi,\xi}=0~~.
\ee
Using previous results, this implies:
\paragraph{Theorem.} Giving $s$ solutions $\xi_1,\ldots,\xi_s$ of \eqref{CGK} which are 
$\cB$-orthonormal everywhere is equivalent to giving $s$ globally-defined endomorphisms 
${\cal O}_1,\ldots,{\cal O}_s\in \Gamma(M,\End(S))$ which satisfy \eqref{ndiagonal} as well as the conditions: 
\ben
\label{Efcdiagonal}
\boxed{D_m^\ad({\cal O}_i)=Q\circ {\cal O}_i=0~~,~~\forall i=1\ldots s~~.}
\een
Furthermore, a solution $({\cal O}_i)_{i=1\ldots s}$ of \eqref{ndiagonal} determines the
 corresponding everywhere $\cB$-orthonormal system of sections of $S$ via the conditions: 
\be
{\cal O}_i=E_{\xi_i,\xi_i}~~,
\ee
up to {\em independent} ambiguities of the form: 
\be
\xi_i\rightarrow -\xi_i~~.
\ee
\noindent Since by the argument recalled above any system of independent solutions can be replaced 
(upon making linear combinations with coefficients from $\cinf$) with a system of 
solutions which are $\cB$-orthonormal everywhere, this result gives a complete characterization of 
nontrivial solutions to the problem \eqref{CGK}.

\section{Application of the reconstruction theorem to spinors and the translation to forms}

We shall present in what follows the implications of the reconstruction theorem to our approach 
through geometric algebra developed in \cite{ga1,ga2,gf}, which uses an isomorphic 
realization of the Clifford bundle ${\rm Cl}(T^*M)$ of $T^* M$ as the Kahler-Atiyah bundle 
$(\wedge T^*M, \diamond)$, where the so-called geometric product $\diamond:\wedge T^*M\times \wedge T^*M\rightarrow \wedge T^*M$ 
is an associative (but non-commutative)
fiberwise composition which makes the exterior bundle into a bundle of unital associative algebras. 

Let $(M,g)$ be a pseudo-Riemannian manifold and $S$ be a pin bundle on $M$, with underlying pin representation
$\gamma:(\wedge T^\ast M,\diamond)\rightarrow (\End(S),\circ)$. Let
$\cB$ be an admissible bilinear form on $S$ (see \cite{gf} for a
detailed discussion), which we assume to be a fiberwise scalar
product.  We also assume that the signature $(p,q)$ of $g$ is such that we are
in the normal case, i.e. the Schur algebra of $\gamma$ equals the base
field $\R$.

With the sub-bundle $\wedge^\gamma T^\ast M$ of $\wedge
T^\ast M$ defined as in \cite{ga1}, the restriction of $\gamma$ gives
an isomorphism of bundles of algebras from $\wedge^\gamma T^\ast M$ to
$\End(S)$, whose inverse we denote by $\gamma^{-1}:\End(S)\rightarrow
\wedge ^\gamma T^\ast M$ (we shall use the same notation for the
induced map on sections). As in \cite{ga1}, we let $\check{T}\eqdef
\gamma^{-1}(T)\in \Omega ^\gamma(M)\eqdef \Gamma(M, \wedge^\gamma T^\ast
M)$ denote the `vertical dequantization' of any globally-defined
endomorphism $T\in \Gamma(M,\End(S))$.

Consider a system of global endomorphisms ${\cal O}_{ij}\in \Gamma(M,\End(S))$ as before, where $i,j=1\ldots s$. 
Applying $\gamma^{-1}$ and using the results of \cite{ga1} we find that \eqref{S} is {\em equivalent} with the 
following system of equations for the inhomogeneous forms $\check{\cal O}_{ij}\eqdef
\gamma^{-1}({\cal O}_{ij})\in \Omega^\gamma(M)$:
\ben
\label{SCheck}
\begin{split}
& \check{\cal O}_{ij}\diamond \check{\cal O}_{kl}=\cS(\check{\cal O}_{kj}) \check{\cal O}_{il}~~,~~\forall i,j,k,l=1\ldots s\\
& \tau_{\cB}(\check{\cal O}_{ij}) = \check{\cal O}_{ji}~~,~~\forall i,j=1\ldots s~~,
\end{split}
\een
while \eqref{diagonal} are {\em equivalent} with the following decoupled systems for the `diagonal' components 
$\check{\cal O}_i\eqdef \check{\cal O}_{ii}$:
\ben
\label{diagonalCheck}
\boxed{
\begin{split}
& \check{\cal O}_i\diamond \check{\cal O}_i= \cS(\check{\cal O}_i) \check{\cal O}_i~~\\
& \tau_\cB(\check{\cal O}_i) = \check{\cal O}_i~~.
\end{split}
}~~,
\een
Here, $\tau_\cB$  is the {\em modified reversion} defined through:
\ben
\label{ModifRev}
\tau_\cB\eqdef \tau\circ \pi^{\frac{1-\epsilon_\cB}{2}} = \twopartdef{\tau~,~}{\epsilon_\cB=+1}{\tau\circ \pi~,~}{\epsilon_\cB=-1}~~,
\een
with $\epsilon_\cB$ discussed in loc. cit., while $\cS:\Omega^\gamma(M)\rightarrow \cinf$ is the 
trace on the reduced \KA algebra $(\Omega^\gamma(M),\diamond)$ defined in \cite{ga1}, 
\ben
\label{TraceDef}
\cS(\omega)\eqdef \omega^{(0)} N_{p,q} \rk S~~,
\een
where $\omega^{(0)}\in \cinf$ denotes the rank zero component of $\omega$ and $N_{p,q}$ equals $1$ or $2$ according to
whether the fiberwise representation $\gamma$ is faithful or not. One has \cite{ga1}:
\be
\cS(\check{T})=\tr(T)~~,~~\forall T\in \Gamma(M,\End(S))~~
\ee
The geometric product $\diamond$ defined in \cite{ga1} can be viewed
as a deformation of the wedge product parameterized by the metric $g$,
which reduces to the latter in the limit $g\rightarrow \infty$.

\

For systems of everywhere orthonormal spinors, the previous results imply:
\paragraph{Proposition.} Giving $s$ smooth global sections $\xi_1,\ldots,\xi_s\in \Gamma(M,S)$ which are
everywhere $\cB$-orthonormal amounts to giving $s$ globally-defined
smooth forms $\check{{\cal O}}_1,\ldots,\check{{\cal O}}_s\in \Omega^\gamma(M)$
specified up to independent signs (i.e. up to independent ambiguities
of the form $\check{{\cal O}}_i\rightarrow -\check{{\cal O}}_i$) such that the
following system of relations is satisfied, where $i,j=1\ldots s$:
\beqan
\label{System}
\boxed{
\begin{split}
&\check{{\cal O}}_i\diamond \check{\cal O}_i=\check{\cal O}_i~~,\\
&\tau_\cB(\check{\cal O}_i)=\check{\cal O}_i~~,\\
& \cS(\check{\cal O}_i)=1_M~~,\\
&\cS(\check{\cal O}_i\diamond \check{\cal O}_j)=0_M~~\mathrm{for}~~i < j~~.
\end{split}}
\eeqan
Furthermore, a solution of this system determines the corresponding sections $\xi_i$ 
through the relations: 
\be
\check{\cal O}_i=\check E_{\xi_i,\xi_i}
\ee
up to {\em independent} ambiguities of the type: 
\be
\xi_i\rightarrow -\xi_i~~,
\ee
where the inhomogeneous forms $\check E^{(k)}_{\xi_i,\xi_j}$ were introduced and discussed in \cite{ga1,ga2,gf}.
\noindent To gain some understanding of the content of these conditions, 
consider the rank expansions: 
\be
\check{\cal O}_j=\sum_{k=0}^d \check{\cal O}_j^{(k)} ~~\mathrm{with}~~\check{\cal O}_j^{(k)}
\in \Omega^k(M)~~.
\ee
Using definition \eqref{ModifRev} of $\tau_\cB$ shows that the
second equations in \eqref{System} amount to the condition that
only those ranks $k\in \{0,\ldots,d\}$ which satisfy the following condition
need to be considered in these rank expansions (i.e., we have $E_j^{(k)}=0$ for
any $k$ which does not satisfy this condition):
\ben
\label{TauCond}
k(k-\epsilon_\cB)\equiv_4 0~~
\een
i.e. $k\equiv_4 0,\epsilon_\cB$.
Using \eqref{TraceDef} shows that the third equations in \eqref{System} amount to:
\ben
\label{TraceCond}
\check{\cal O}_j^{(0)}=\frac{1}{N_{p,q}\rk S}~~.
\een
Relation \eqref{TraceCond} suggests the rescaling 
$\bcO^{(k)}\eqdef (N_{p,q}\rk S)\check{\cal O}^{(k)}\in \Omega^k(M)$, so that:
\ben
\check{\cal O}_i=\frac{1}{N_{p,q}\rk S}\sum_{k=0}^d{\bcO^{(k)}_i}~~.
\een
Then \eqref{TraceCond} amounts to the condition $\bcO^{(0)}_i =1$ and hence 
the second and third rows of \eqref{System} are equivalent with the statement that $\check{E}_i$ have the following rank expansions: 
\ben
\label{RescaledExpansion}
\check{\cal O}_i=\frac{1}{N_{p,q}\rk S}\left[1_M+\sum_{\tiny \begin{array}{c}k=1\\
 k(k-\epsilon_\cB)\equiv_4 0\end{array}}^d {\bcO^{(k)}}\right]~~.
\een

\paragraph{Constrained generalized Killing spinors.}
Consider the situation in which $S$ is endowed with a $\cB$-compatible 
connection $D_m$ and with a global 
endomorphism $Q\in \Gamma(M,\End(S))$. Then solutions of 
\eqref{CGK} will be called {\em constrained generalized Killing spinors} (CGKS). 
As shown in \cite{ga1}, 
the connection $D_m$ induces 
a derivation $\check{D}_m$ of the reduced \KA algebra $\Omega^\gamma(M)$ while $Q$ 
induces an element 
$\check{Q}\eqdef \gamma^{-1}(Q)$ of this algebra. Furthermore, 
we have $\check{D_m(T)}=\check{D}_m(\check{T})$ and 
$\check{Q\circ T}=\check{Q}\diamond\check{T}$ for all $T\in \Gamma(M,\End(S))$. 
Hence the results of previous subsection give the following theorem, 
which gives a precise mathematical encoding 
of the ``method of bilinears'' \cite{Tod} in the situation at hand: 

\paragraph{Theorem.} Giving $s$ globally-defined smooth spinors 
$\xi_1,\ldots,\xi_s$ which satisfy \eqref{CGK} 
and which are $\cB$-orthonormal everywhere is equivalent to giving $s$ 
globally-defined
forms $\check{\cal O}_1,\ldots,\check{\cal O}_s\in \Omega^\gamma(M)$ which 
satisfy \eqref{System} as well as the conditions: 
\ben
\label{GKSdiagonal}
\boxed{\check{D}_m^\ad(\check{\cal O}_i)=\check{Q}\diamond \check{\cal O}_i=0_M}~~,
~~\forall i=1\ldots s~~.
\een
Furthermore, a solution $(\check{\cal O}_i)_{i=1\ldots s}$ of \eqref{System} 
determines the corresponding 
everywhere $\cB$-orthonormal system of sections of $S$ via the conditions: 
\be
\check{\cal O}_i=\check{E}_{\xi_i,\xi_i}~~,
\ee
up to {\em independent} ambiguities of the form: 
\be
\xi_i\rightarrow -\xi_i~~.
\ee

\section{Application to compactifications of M-theory on 8-manifolds}
\label{sec:4}

Since for these backgrounds $p-q\equiv_8 0$, we are in the normal simple case, so
$\Omega^\gamma(M)=\Omega(M)$.  For these values of $p$ and $q$ (namely
$p=8$ and $q=0$), one has (up to rescalings by smooth nowhere
vanishing real-valued functions defined on $M$) two admissible
pairings $\cB_\pm$ on $S$ (see \cite{gf}), both of which are
symmetric and have the types $\epsilon_{\cB_\pm}=\pm 1$. Since any
choice of admissible pairing leads to the same result, we choose to
work with $\cB\eqdef \cB_+$ without loss of generality.  Then
$\tau_\cB$ coincides with the canonical reversion $\tau$ of the \KA
algebra of $(M,g)$, which is defined through:
\ben
\label{TauCond2}
\tau(\omega)\eqdef (-1)^{\frac{k(k-1)}{2}}\omega~~,~~\forall \omega\in \Omega^k(M)~~.
\een
Upon rescaling by a smooth function, we can in fact take $\cB$ to be a
fiberwise scalar product on $S$ and denote the corresponding norm
through $||~||$.

\subsection{The case $s=1$}

This was studied in \cite{MartelliSparks, Tsimpis} and more completely in \cite{ga1} using 
geometric algebra techniques. Below we recall the main results 
as an illustration of our reconstruction theorem and as a preparation for the case ${\cal N}=2$. 

As stated in Section \ref{sec:2}, the condition that $S$ admits a 
nowhere-zero section which is everywhere 
of rank one is {\em equivalent} with the condition that \eqref{System} admits at least one 
solution $\check{\cal O}\in \Omega(M)=\Omega^\gamma(M)$.
Since $N_{8,0}=1$ and $\rk S=16$, the general definition \eqref{TraceDef} becomes:
\be
\cS(\omega)=16 \omega^{(0)}~~,~~\forall \omega\in \Omega(M)~~.
\ee
The second relation of \eqref{System} means that the rank expansion of $\check{\cal O}$ contains only 
forms of ranks $k=0,1,4,5$ and $8$.
This case was studied in \cite{MartelliSparks, Tsimpis} and more completely in our work \cite{ga1}
using K\"{a}hler-Atiyah algebra techniques. 
The non-quadratic relations in the reconstruction theorem applied to this case 
amount to the statement that $\check{\cal O}\in \Omega(M)$ expands as
\ben
\label{Qexp}
\check{\cal O}=\frac{1}{16}(1+ V+\Phi+Z+b\nu)~~,
\een
where $b\in \cinf$, $V\in \Omega^1(M), \Phi\in \Omega^4(M)$, $Z\in \Omega^5(M)$ and $\nu$ 
is the canonical volume form of $(M,g)$. 
In \cite{ga1}, we used the
normalization $||\xi||=\sqrt{2}$ rather than $||\xi||=1$ as we do in
this paper.  As a consequence, the quantities $a, b, {\bar K}, Y$ and
${\bar Z}$ used in that paper correspond to what we call $a, b, V,
\Phi$ and $Z$ in this paper up to a factor of 2. 
Defining $\varphi\eqdef \ast Z\in \Omega^3(M)$, one finds the following relations which hold globally on $M$: 
\ben
\label{Phi}
(1 \mp b)\Phi^\pm= \pm (V\wedge \varphi)^\pm~~.
\een
as well as:
\ben
\label{SolMS}
||V||^2=1-b^2~~,~~||\varphi||=\sqrt{7} (1-b^2) ~~,~~\iota_V\varphi=0~~.
\een
One also finds the globally valid relation:
\ben
\label{G2rel}
||V||^2 ||\iota_{x\wedge y}\varphi||^2=||x\wedge y\wedge V ||^2~,~\forall x,y\in \Omega^1(M) ~~.
\een
The reconstruction theorem tells us that imposing 
\ben
\label{Fierz1}
\check{\cal O}\diamond\check{\cal O}=\check{\cal O}~
\een 
{\em guarantees} that 
$\check{\cal O}=\check{E}_{\xi,\xi}$ for
some globally-defined normalized spinor $\xi\in \Gamma(M,S)$ which 
is determined up to sign by this condition. This recovers the Fierz identities of 
Martelli-Sparks purely from \KA algebra relations. 
The CGKS equations in \KA form
\be
\check{D}_m^\ad(\check{\cal O})=\check{Q}\diamond \check{\cal O}=0_M~
\ee 
recover the remaining results in \cite{MartelliSparks, Tsimpis}. 

The geometry can be summarized by saying that $M$ admits a singular
(Stefan-Sussmann) foliation whose leaves of codimension one carry a
longitudinal $G_2$ structure and whose leaves of codimension zero
carry (positively or negatively oriented) $\Spin(7)$ structures.  The
codimension zero leaves are subsets on $M$ along which one of the two
Weyl components of $\xi$ vanishes (which one of them vanishes
determines the orientation of the corresponding $\Spin(7)$ structure).
In general, the structure group of $M$ does not globally reduce, which
is why one needs the theory of Stefan-Sussmann foliations rather than
regular foliation theory.  The torsion coefficients of these
longitudinal $G$-structures can be computed explicitly using a
combination of geometric algebra methods with various properties of
$G_2$ and $\Spin(7)$ structures given in \cite{KarigiannisThesis,
  KarigiannisFlows}. We  can also extract the
complete expressions for the background fluxes, which have not yet
been given in the literature.

Let $\cD_V\eqdef \ker V=\{X\in \cX(M)|V(X)=0\}=\{X\in
\cX(M)|g(V^\sharp, X)=0\}\subset \cX(M)$ be the singular
(Stefan-Sussmann) distribution defined by the orthogonal complement of
the vector field $V^{\sharp}$. Its tangent
fiber $\cD_V(p)$ at
a point $p\in M$ has rank seven when $V(p)\neq 0$ and rank eight when
$V(p)=0$, the last condition defining a closed subset of $M$ which in turn is a disjoint union of 
subsets $M_V^+$ and $M_V^-$ consisting of those points of $M$ where one of the two Weyl 
components of $\xi$ vanishes. 

Conditions \eqref{SolMS} and \eqref{G2rel} mean that, on the $G_2$ locus $U_V\eqdef M\setminus(M_V^+\sqcup M_V^-)$, the 
3-form $\frac{1}{||V||}\varphi$ 
is longitudinal to $\cD_V$ and its restriction to 
$\cD_V$ is the canonically normalized $3$-form of a $G_2$ structure on $\cD_V$, 
which is compatible with the metric given by the restriction of $g$ to $\cD_V$. On the locus $M_\pm$, the singular distribution $\cD_V$
coincides with the restriction of the tangent space and one finds that the retstriction of the four-forms $\Phi^\pm$ satisfy the defining 
conditions of a $\Spin_\pm(7)$ structure. 
The Majorana spinor $\xi$ is determined (up to sign) by the smooth function 
$b\in \cC^\infty(M,[-1,1])$ and by the following data: 

(a) the Frobenius distribution $D_V|_{U_V}$ together with its $G_2$ structure

(b) the $\Spin(7)$ structures on the two topological bundles $TM|_{M_V^{(\pm)}}$ (whose orientations relate to that of $TM$ as explained above). 

\paragraph{Remark.} A very particular case in which $b$ is everywhere degenerate is the case 
$b=+1$ everywhere (the case $b=-1$ everywhere is very similar) 
i.e. $\xi\in \Gamma(M,S_+)$ (then $\dd_p b=0$ for all $p\in M$). In this case, 
$V$ vanishes identically and $D_V=TM$ has constant rank 
equal to eight, thus all points of $M$ are regular for $D_V$.    
Relations \eqref{SolMS} give $\varphi=0$ while \eqref{G2rel} and \eqref{Phi} are 
trivially satisfied. Relations
\be
||\Phi||^2=14~~,~~\ast \Phi=\Phi~~,~~\Phi \btu_2\Phi=-14 \Phi~~
\ee 
reflect the fact that $\xi$ (equivalently, $\Phi$) defines a canonically-normalized 
$\Spin(7)$ structure on $M$ which is compatible with the metric.

\subsection{The case $s=2$}

As previously stated, here and in our previous work \cite{ga2}, the internal parts of our supersymmetry generators are not assumed to have 
definite chirality\footnote {This was also attempted in \cite{Palti} for the same compactifications.}, 
which is a surprisingly nontrivial generalization of what 
was previously studied in the literature \cite{BB}, complicating computations quite drastically. 

In this case we have two spinors in eight dimensions, and the results 
of Sections \ref{sec:2} and \ref{sec:3} show that giving two globally-defined 
spinors on $M$ which are everywhere $\cB$-orthonormal is {\em equivalent} to giving 
a solution $(\check{\cal O}_1,\check{\cal O}_2)$ of 
the following system of equations:
\beqan
\label{SystN2D8}
&&\check{\cal O}_1\diamond \check{\cal O}_1=\check{\cal O}_1~~,
~~\check{\cal O}_2\diamond \check{\cal O}_2=\check{\cal O}_2\\
&&\tau(\check{\cal O}_1)=\check{\cal O}_1~~,~~\tau(\check{\cal O}_2)=\check{\cal O}_2\\
&&\cS(\check{\cal O}_1)=\cS(\check{\cal O}_2)=1_M\\
&&\cS(\check{\cal O}_1\diamond \check{\cal O}_2)=0_M~~~
\eeqan
The first three rows form independent systems for $\check{\cal O}_1$
and $\check{\cal O}_2$ of the type studied before (they characterize the
existence of everywhere $\cB$-normalized global sections $\xi_1,
\xi_2$ of $S$) while the last relation enforces $\cB$-orthogonality
everywhere of $\xi_1$ and $\xi_2$ (and, in particular, linear
independence of these spinors everywhere). Using the results 
for the case $s=1$, we therefore know that the first three rows of
\eqref{SystN2D8} are solved by:
\ben
\label{Eexpi}
\check{\cal O}_i=\frac{1}{16}(1+ V_i+\Phi_i+Z_i+b_i\nu)~~,
\een
where $V_i\in \Omega^1(M),\Phi_i\in \Omega^4(M), Z_i\in \Omega^5(M)$
and $b_i\in \cinf$, for $i=1$ and $i=2$.

Applying the reconstruction theorem to this case shows 
that the full system of Fierz relations \eqref{SystN2D8} plus the GKS 
constraint \eqref{GKSdiagonal} (differential and algebraic) 
\be
\check{D}_m^\ad(\check{\cal O}_i)=\check{Q}\diamond \check{\cal O}_i=0_M~~,~~i=1,2
\ee 
is equivalent with two copies of the system found in the $s=1$ 
case plus a single  quadratic constraint which couples these 
two systems (the 4th relation in \eqref{SystN2D8}). This gives a drastic 
simplification of the system of relations that we extracted 
in previous work and provides a way to describe the geometry in terms of 
certain foliations. Namely, we find that, on the `generic locus', $M$ carries a {\em codimension three} foliation endowed 
with a five dimensional $\SU(2)$ structure in the sense of Conti and Salamon \cite{ContiSalamon}. Once again, one needs the 
theory of Stefan-Sussmann foliations, which also produces other loci beyond the generic locus, carrying lower codimensional 
foliations endowed with different longitudinal $G$-structures. 

The compactification space $M$ becomes a Calabi-Yau fourfold when $s=2$ and 
$K(D,Q)\subset \Gamma(M,S_+)$ or 
$K(D,Q)\subset \Gamma(M,S_-)$, in which case $\kappa=0$ and 
the CGK equations amount to the conditions 
that $f=d\Delta^{-3/2}$ and $F$ a 
primitive $(2,2)$ form \cite{BB}. 
However, there is no reason to require that $K(D,Q)$ consists of 
Majorana-Weyl spinors! Therefore, our general solutions can be 
called ``generalized Calabi-Yau fourfolds''. 

They are of interest for generalizing (and thus solving the problem 
of $G$-flux) in F-theory by considering an F-theory limit which should 
make sense when the leaves of our foliations 
admit $T^2$ fibrations.

\section{Generalization and conclusions}
\label{sec:5}

Using the reconstruction theorem, one can similarly characterize supersymmetric 
backgrounds of this type which preserve more than two supersymmetries, 
since the reconstruction theorem allows one to reduce the case of 
$s$ CGK spinors to $s$ copies of a single CGK spinor plus a set of 
$\frac{s(s-1)}{2}$ quadratic algebraic constraints, thus allowing a 
systematic study for $s>1$. In general, one finds a description though Stefan-Sussmann foliations carrying various 
longitudinal $G$-structures.

\subsection*{Acknowledgements} The work of E.M.B. was supported by the strategic grant 
POSDRU/159/1.5/S/133255, Project ID 133255 (2014), co-financed by the European Social 
Fund within the Sectorial Operational Program Human Resources Development 2007--2013, 
and also by CNCS-UEFISCDI through Contract Idei-PCE, No. 121/2011, while C.I.L. 
acknowledges support from the Research Center Program of the Institute 
for Basic Science (IBS) in Korea within the grant CA1205-01.

\end{document}